\documentstyle[fleqn,iopfts,12pt]{ioplppt}
\jl{1}
\begin{document}
\title{Reply to the Comment on ``On the uncertainty relations and squeezed 
states for the quantum mechanics on a circle''}
\author{K Kowalski and J Rembieli\'nski}
\address{Department of Theoretical Physics, University
of \L\'od\'z, ul.\ Pomorska 149/153,\\ 90-236 \L\'od\'z,
Poland}
\section*{}
In the preceding Comment [1] Trifonov disputes our uncertainty
relations for a quantum particle on a circle recently proposed in
[2] such that
\begin{equation}
\Delta^2(\hat\varphi)+\Delta^2(\hat J)\ge1,
\end{equation}
where $\Delta^2(\hat\varphi)$ and $\Delta^2(\hat J)$ are measures of
the uncertainty of the position and angular momentum, respectively.
He states that (i) the quantity $\Delta^2(\hat\varphi)$ introduced
in [2] representing the uncertainty of the angle is not a proper
measure of the position uncertainty and therefore the proposed
inequality (1) can hardly be qualified as a relevant uncertainty
relations on a circle; and that (ii) the most suitable uncertainty
relations on a circle are those based on the Gram-Robertson matrix
[3].  We disagree with both points.

(i)\quad We recall that Trifonov [1] provides an example of the
state which can be regarded as a counterpart of the Schr\"odinger
cat state in the case of the circular motion, such that the
corresponding wave packet seems to be worse localized than that
referring to the coherent state for a quantum particle on a circle,
despite the fact that the uncertainty $\Delta^2(\hat\varphi)$ in the
Schr\"odinger cat state is lesser than in the coherent state.  In
our opinion the discussion of the uncertainty relations cannot be
confined, as done by Trifonov [1], to the localization in the
configuration space but it must take into consideration the
localization in the {\em phase space}.  Reasoning analogously as
Trifonov one could provide the following ``proof'' of the irrelevance
of the uncertainty relations for the sum of variances of the position 
and momentum of a particle on a real line implied by the standard
Heisenberg uncertainty relations, of the form
\begin{equation}
\Delta^2\hat x +\Delta^2\hat p\ge1,
\end{equation}
where we set $\hbar=1$.

Consider the wavefunctions such that [4]
\begin{eqnarray}
\psi(x) &=& 
\cases{1/\sqrt{L}, &for $-L/2<x<L/2$,\cr
0, &for $-L/2>x>L/2$,}\\
\phi(x) &=& 
\cases{\sqrt{2/L}, &for $-L/2<x<-L/4$,\cr
0, &for $-L/4<x<L/4$,\cr
\sqrt{2/L}, &for $L/4<x<L/2$,\cr
0, &for $-L/2>x>L/2$,}
\end{eqnarray}
where $L>0$.  As one can see the state $|\psi\rangle$ is much worse 
localized on the interval $|x|<L/2$, than the state $|\phi\rangle$.  
In fact, we know that in the state $|\phi\rangle$ the particle is not in the
region $|x|<L/4$.  However, when we calculate the variances we get
\begin{equation}
\Delta^2_\psi \hat x = \frac{L^2}{12},\qquad \Delta^2_\phi \hat x =
\frac{7}{4}\frac{L^2}{12}.
\end{equation}
Thus it turns out that that the variance in the state $|\psi\rangle$
is considerably lesser than in the state $|\phi\rangle$.  Therefore,
concluding the ``proof" ---
the variance is not a proper measure of the position uncertainty and
the Heisenberg uncertainty relations could hardly be qualified as a
relevant uncertainty relations on a line.

Finally, we would like to
stress that the motivation for the usage in [2] the denomination
``squeezed states'' was only the formal similarity of generation of
these states and the standard squeezed states.  In particular,
neither any squeezing property was discussed nor any definition
provided in [2] like ``The quantity $\tilde \Delta^2(\hat\varphi)$ is
called squeezed if it is less than 1/2'' as erroneously indicated in
[1].  Moreover, the problems were reported in [2] with the physical
interpretation of the parameter $s$ labelling the squeezed states
for the quantum mechanics on a circle.

(ii)\quad The uncertainty relations on a circle proposed by Trifonov
[1] utilize (generalized) variances of the angle.  We share the opinion of
Bia{\l}ynicki-Birula {\em et al\/} [5] that: ``Second moment or variance
\ldots This is a naive extension of the mathematical formulation of
uncertainty which is used for Heisenberg's position-momentum
uncertainty relation.  The main drawback of this measure of
uncertainty is that we evaluate the averages of non-periodic
function, such as $\varphi$ or $\varphi^2$, with a periodic
distribution function.  Consequently this measure can assume
completely arbitrary values depending on the origin of the phase
integration, that is on the coordinatization of the unit circle.''
In fact, since there is no distinguished point on a circle,
therefore it is clear that the uncertainty of the position of a
quantum particle should depend solely on its state and not the choice
of the particular point on a circle.  Evidently, this is not the
case when we apply the standard variance.  Namely, we find
\begin{equation}
\Delta ^2_\lambda\hat \varphi  - \Delta ^2_0\hat \varphi =
2\int\limits_{0}^{\lambda}(\varphi +\pi -\langle
\varphi\rangle_0)|f(\varphi)|^2d\varphi
-\left(\int\limits_{0}^{\lambda}|f(\varphi)|^2d\varphi\right)^2,
\end{equation}
where
\begin{equation}
\Delta ^2_\lambda\hat \varphi = \langle \hat\varphi^2\rangle_\lambda -
\langle \hat\varphi\rangle^2_\lambda =
\frac{1}{2\pi}\int\limits_{\lambda}^{2\pi+\lambda}\varphi^2|f(\varphi)|^2
d\varphi-\left(\frac{1}{2\pi}\int\limits_{\lambda}^{2\pi+\lambda}\varphi
|f(\varphi)|^2d\varphi\right)^2,
\end{equation}
and the normalized wave packet $f(\varphi)$ is a $2\pi$-periodic function, 
i.e.\ $f(\varphi +2\pi)=f(\varphi)$.  For an easy illustration of the 
dependence of the variance $\Delta ^2_\lambda\hat
\varphi$ on the origin of integration $\lambda$ we now discuss the
case of the normalized wave packet of the form
\begin{equation}
f(\varphi) = \sqrt{\frac{2\pi}{\epsilon}}\chi_{[0,\epsilon]}(\varphi),
\end{equation}
where $\chi_{[0,\epsilon]}(\varphi)$ is the characteristic function
of the interval $[0,\epsilon]$ and $\epsilon\in(0,2\pi)$.  Of
course, the wave packet (8) can be made $2\pi$-periodic by taking
the interval $[0,\epsilon]$ modulo $2\pi$.  Now the straightforward
calculation shows that the difference of variances (6) for the wave
packet (8) is
\begin{eqnarray}
\Delta ^2_\lambda\hat \varphi  - \Delta ^2_0\hat \varphi&=& \left\{
\begin{array}{ll}
0 & \hbox{for }\epsilon\le\lambda\\
\frac{2\pi}{\epsilon}\lambda
\left[\left(1-\frac{2\pi}{\epsilon}\right)\lambda
+2\left(\pi-\frac{\epsilon}{2}\right)\right] & \hbox{for
}\epsilon>\lambda .
\end{array}
\right.
\end{eqnarray}
Thus, as expected, the difference of variances (6) depends in general
on the origin of integration $\lambda$.
We would like to point out that in a sense Trifonov seems to recognize 
the discussed flaw of the standard variance since he suggests in [1] that ``the
mean values $\langle\varphi\rangle$, $\langle\varphi^2\rangle$
should be calculated by integration from $\varphi_0-\pi$ to
$\varphi_0+\pi$, where $\varphi_0$ is the centre of the wave packet
(i.e.\ $\varphi_0$ is the most probable value of $\varphi$)''.  In
our opinion such solution of the problem which introduces
the {\em definition\/} of average values depending on the particular state
of the system can hardly be called satisfactory.  Another evidence
that the variance utilized by Trifonov [1] can hardly be qualified
as a relevant uncertainty of the position on a circle is the ill
behaviour of the expectation value $\langle\hat\varphi(t)\rangle$ in
the case of the free evolution of the coherent states.  Namely, it
turns out that $\langle\hat\varphi(t)\rangle$ takes the values only
from subset of the circle $[0,2\pi)$ [6].

We would like to stress that our measure of the uncertainty of the
position of a quantum particle on a circle given by [2]
\begin{equation}
\Delta^2(\hat\varphi) = -\frac{1}{4}{\rm ln}|\langle U^2\rangle|^2,
\end{equation}
where $U=\exp({\rm i}\hat\varphi)$, has correct behaviour and does
not depend on the origin of the integration.  Indeed, we have
\begin{equation}
\Delta^2_\lambda(\hat\varphi) = \Delta^2_0(\hat\varphi),
\end{equation}
where
\begin{equation}
\Delta^2_\lambda(\hat\varphi) = -\frac{1}{4}{\rm ln}|\langle
U^2\rangle_\lambda|^2,
\end{equation}
and
\begin{equation}
\langle U^2\rangle_\lambda = \frac{1}{2\pi}\int\limits_{\lambda}^{2\pi+\lambda}
e^{2{\rm i}\varphi}|f(\varphi)|^2d\varphi,
\end{equation}
following immediately from
\begin{equation}
\langle U^2\rangle_\lambda = \langle U^2\rangle_0.
\end{equation}

We remark that an interesting observation of Trifonov [1]
is that the uncertainty relations are minimized by the
Schr\"odinger-cat like states mentioned earlier.  Therefore, in
oposition to the standard coherent states, the coherent states for
the quantum mechanics on a circle are not uniquely determined, up to
a unitary transformation, by the requirement of the saturation of
the uncertainty relations (1).  Nevertheless, the topology of the circle 
is completely different from the topology of the real line and it seems 
plausible that the coherent states for a quantum particle on a circle may 
have some properties different from those of the standard coherent states 
referring to the case of the real line.  We also point out that
Trifonov have not provided in [1] any example of states violating
the inequality (1).

Finally, we would like to comment on the note added to proof [1],
that ``\ldots coherent states have been introduced (in more general
notations) by S de Bievre and J Gonzales in 1993 [2]''.  First of all, 
we introduced in [7] the
coherent states for a quantum particle on a circle as a solution of some 
eigenvalue equation independently of the treatment of Gonzales 
{\em et al\/} [8] who applied the Weil-Brezin-Zak transform.  We
stress that our approach based on a polar decomposition of an
operator defining via the eigenvalue equation the coherent states
enabled us to construct the coherent states for the quantum
mechanics on a sphere [9].  We remark that both coherent states for
a quantum particle on a circle and on a sphere are concrete
realizations of the general mathematical scheme of construction of
the Bargmann spaces introduced in [10,11] (see also recent work [12]).
The ``more general notations'' mentioned by Trifonov are connected
with the fact that the coherent states utilized by Gonzales {\em et
al\/} [8] are labelled by some parameter which can be avoided
by demanding the time-reversal invariance [7] which leads precisely to
the coherent states introduced in [7].

\end{document}